\documentclass{aa}
\usepackage{graphicx}
\usepackage{times}
\newcommand{\ab}[1]{\mbox{Fig.\,\ref{#1}}}
\newcommand{\ang}{\mbox{\rm \AA}}

\newcommand{\ea}{et\,al.}

\newcounter{Rco}

\newcommand{\Ionst}[1]{\setcounter{Rco}{#1}\Roman{Rco}}
\newcommand{\Ion}[2]{\mbox{#1\ {\scriptsize\Ionst{#2}}}}

\newcommand{\Ionw}[3]{\mbox{#1\ {\scriptsize\Ionst{#2}}~$\lambda\,#3$\AA}}
\newcommand{\Ionww}[3]{\mbox{#1\ {\scriptsize\Ionst{#2}}~$\lambda\lambda\,#3$\AA}}

\newcommand{\kK}{\mbox{\rm kK}}
\newcommand{\logg}{\mbox{$\log g$}}
\newcommand{\loggw}[1]{\mbox{$\log g\hspace{-0.5mm} =\hspace{-0.5mm}  #1$}}
\newcommand{\mc}[3]{\multicolumn{#1}{#2}{#3}}

\newcommand{\ratio}[2]{\mbox{$n_{\rm #1}/n_{\rm #2}$}}
\newcommand{\ratiow}[3]{\mbox{$n_{\rm #1}/n_{\rm #2}\hspace{-0.5mm} = \hspace{-0.5mm} #3$}}
\newcommand{\sA}[1]{\mbox{(Fig.\,\ref{#1})}}

\newcommand{\sga}{\mbox{$\stackrel{>}{{\mbox{\tiny $\sim$}}}$}}

\newcommand{\sla}{\mbox{$\stackrel{<}{{\mbox{\tiny $\sim$}}}$}}

\newcommand{\spm}{\mbox{\raisebox{0.20em}{{\tiny \hspace{0.2mm}\mbox{$\pm$}\hspace{0.2mm}}}}}
\newcommand{\sK}[1]{\mbox{(Sect.\,\ref{#1})}}
\newcommand{\se}[1]{\mbox{Sect.\,\ref{#1}}}
\newcommand{\sT}[1]{\mbox{(Tab.\,\ref{#1})}}
\newcommand{\ta}[1]{\mbox{Tab.\,\ref{#1}}}
\newcommand{\Teff}{\mbox{$T_\mathrm{eff}$}}
\newcommand{\Teffw}[1]{\mbox{$\Teff\hspace{-0.5mm} =\hspace{-0.5mm} #1 \mathrm{kK}$}}

\newcommand{\AuA}[2]{A\&A #1, #2 }

\newcommand{\AuAS}[2]{A\&AS #1, #2 }
\newcommand{\AJ}[2]{AJ #1, #2 }

\newcommand{\APJ}[2]{ApJ #1, #2 }
\newcommand{\APJS}[2]{ApJS #1, #2 }

\newcommand{\MNRAS}[2]{MNRAS #1, #2 }
\newcommand{\PASP}[2]{PASP #1, #2 }
\begin{document}
%
%
   \title
         {Spectral analysis of the sdO \object{K\,648}, 
          the exciting star of the planetary nebula \object{Ps\,1} 
          in the globular cluster \object{M\,15} (\object{NGC\,7078})\thanks
              {Based on observations obtained at the German-Spanish Astronomical Center,
               Calar Alto, operated by the Max-Planck-Institut f\"ur Astronomie Heidelberg
               jointly with the Spanish National Commission for Astronomy,
               on data retrieved from the International Ultraviolet Explorer (IUE) Final Archive,               
               on observations made with the Hubble Space Telescope (HST, GO Proposal ID: 3513, PI: Heber),
               and on HST data retrieved from the ST-ECF archive.}
         }
   \author
          {T.\,Rauch\inst{1}
      \and U.\,Heber\thanks{Visiting astronomer, Calar Alto, Spain}\inst{2}
      \and K.\,Werner\inst{1}
          }

   \offprints{T.\,Rauch}
   \mail{rauch@astro.uni-tuebingen.de}
 
   \institute
         {Institut f\"ur Astronomie und Astrophysik, Universit\"at T\"ubingen, D-72076 T\"ubingen, Germany
    \and  Dr.\,Remeis-Sternwarte, Universit\"at Erlangen-N\"urnberg, D-96049 Bamberg, Germany
         }
 
    \date{Received 17 July 2001 / Accepted 26 October 2001}

   \titlerunning{Spectral analysis of \object{K\,648}, the exciting star of \object{Ps\,1} in \object{M\,15}}
   \abstract{
             We present a spectral analysis of the sdO central star \object{K\,648} based on medium-resolution
             optical and high-resolution UV spectra. The photospheric parameters are determined by means
             of state-of-the-art NLTE model atmosphere techniques. 
             We found \Teffw{39\spm 2} and \loggw{3.9\spm 0.2}. The helium (\ratiow{He}{H}{0.08})
             and oxygen (\ratiow{O}{H}{0.001}) abundances are about solar while carbon is enriched by a factor
             of 2.5 (\ratiow{C}{H}{0.001}). 
             Nitrogen (\ratiow{N}{H}{1\cdot 10^{-6}}, [N/H] = -2.0) appears at a sub-solar value.
             However, these metal abundances are much higher than the cluster's metallicity (\object{M\,15}: [Fe/H] = -2.25). 
             The surface composition appears to be a mixture of the original hydrogen-rich material
             and products of helium burning ($3\alpha$ process) which have been mixed up to the surface.
             The abundances of He, C, and N are consistent with the nebular abundance, while O is
             considerably more abundant in the photosphere than in the nebula.
             From a comparison of its position in the $\log \Teff$--$\log g$ plane with evolutionary
             calculations a mass of $0.57^{+0.02}_{-0.01}\,\mathrm{M}_\odot$ and a 
             luminosity of $3810\spm 1200\,\mathrm{L}_\odot$ are deduced.
             Our spectroscopic distance $d = 11.1^{+2.4}_{-2.9}\,\mathrm{kpc}$ is in agreement with the
             distance of \object{M\,15} as determined by Alves \ea\, (2000). 
             From the GHRS spectra we measure a radial velocity of $v_\mathrm{rad} = -130\,\mathrm{km/sec}$.
            }
   \maketitle

\begin{figure*}[ht]
  \centering
  \includegraphics[width=\hsize]{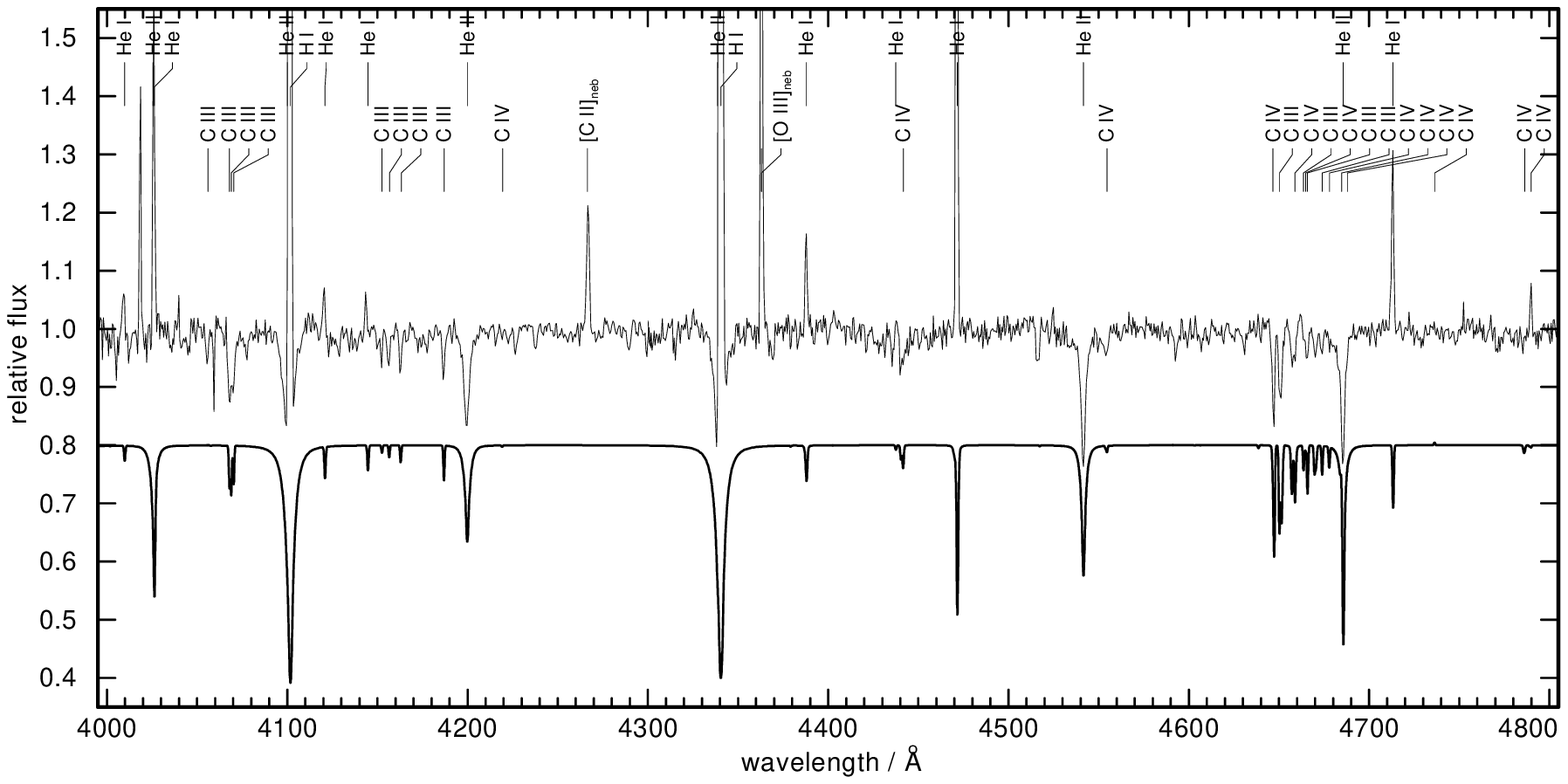}
  \caption[]{Optical spectrum of K\,648 obtained with the TWIN spectrograph (top).
             Positions of H\,{\sc i}, He\,{\sc i-ii}, and C\,{\sc iii-iv} lines are indicated. 
             Note the contamination by nebula emission, e.g. around H\,$\gamma$ and
             \Ionw{He}{1}{4471}. A synthetic spectrum (parameters from 
             \ta{php}) is shown at the bottom 
            }
  \label{twin}
\end{figure*}

\section{Introduction} 
\label{int}

\object{K\,648} ($m_\mathrm{V} = 14.73$, Alves \ea\,2000) was registered by K\"ustner (1921) close to the center 
($\Delta\alpha = +14\arcsec$, $\Delta\delta = +26\arcsec$)
of the galactic globular cluster \object{M\,15}.
A spectrum taken by Pease (1928) shows a continuous O-type spectrum and characteristic emission lines of a 
planetary nebula (\object{Ps\,1}, PN\,G065.0-27.3). Pease derived a radial velocity of 
$v_\mathrm{rad} = -180 \spm 50 \mathrm{km/sec}$
from his spectrum and concluded that \object{K\,648} is probably the first PN discovered in a globular cluster.

Joy (1949) improved the radial-velocity measurement ($v_\mathrm{rad} = -115 \mathrm{km/sec}$) and concluded that, 
without doubt, Ps\,1 is a member of M\,15 ($v_\mathrm{rad} = -107 \mathrm{km/sec}$, Harris 1996) 
because of the close agreement of their radial velocities.

The existence of a PN in \object{M\,15} is unexpected: At a turn-off age of 12 billion years, the most
massive main-sequence stars in \object{M\,15} should have $M_\mathrm{initial} \sla 0.8\,\mathrm{M_\odot}$. 
For such low-mass stars it is
almost impossible to ascend to the AGB and eject a PN. An analysis of the nebula properties by
Adams \ea\ (1984) yields a solar He/H ratio and a slightly higher than solar C/H ratio while the total
N, O, and Ne abundance is less than solar by a factor of 18 \sA{x2h}. The high carbon abundance in the PN was
interpreted by Adams \ea\ as a product of helium burning ($3\alpha$ process) and a subsequent third
dredge-up which has brought the material to the stellar surface. The nebula ejection has taken place then
after this event.

Alves \ea\, (2000) suggest that the progenitor
of \object{K\,648} experienced mass augmentation in a close binary merger and thus, the remnant has
a higher mass than remnants of single stars in \object{M\,15}.

A preliminary analysis of the optical and GHRS spectra \sT{stecf} by means of line-blanketed NLTE model
atmosphere techniques by
Heber \ea\ (1993) revealed \Teffw{37}, \loggw{4.0}, \ratiow{He}{H}{0.5} (by number) and
a three times solar carbon abundance. Since this is much higher than the metallicity of \object{M\,15}
an explanation for the evolutionary history of \object{K\,648} and its PN is difficult.
Bianchi \ea\,(1995, see this paper for a more detailed introduction) presented a first analysis of HST 
data and arrived at \Teffw{35}. Another spectral analysis based on high-resolution Keck spectra was
presented by McCarthy \ea\, (1997). They arrived at \Teffw{43}, \loggw{3.9}, \ratiow{He}{H}{0.08} and a solar
carbon abundance. However, in contrast to Heber \ea\ (1993), NLTE line blanketing had not been taken in
account in these models.
 
Recently a LTE abundance analysis was presented by Bianchi \ea (2001): They used \Teff\ and \logg\
from Heber \ea (1993) and arrived even at a two times higher helium abundance of \ratiow{He}{H}{0.6}, 
carbon is four times solar, oxygen underabundant, and silicon is about solar.

In order to investigate the enigma of \object{K\,648}, we have performed a new NLTE spectral analysis 
based on state-of-the-art metal-line blanketed model atmospheres.

\section{Observations}
\label{obs}

\subsection{Ultraviolet spectra}
\label{uv}

The following spectral analysis is mainly based on HST ``preview'' spectra \sT{stecf} which were retrieved from
the archive operated at ST-ECF.

\begin{table*}[ht]
\caption[]{Summary of the archival HST spectra which are used in this analysis}
\label{stecf}
\begin{tabular}{ccccr@{-}lr@{.}lc}
\hline
\noalign{\smallskip}
image \# & instrument & grating & 
  exposure time / sec & \mc{2}{c}{wavelength range / \AA} & \mc{2}{c}{resolution / \AA} & date \\
\hline
\noalign{\smallskip}
Y1C40105T & FOS  & G130H &                 1\,800 & \hbox{}\hspace{5mm}1153&1605 & \hbox{}\hspace{5mm}0&96 & 93-11-18 \\
Y1C40106T & FOS  & G130H &                 1\,800 &                    1153&1605 &                    0&96 & 93-11-18 \\
Y1C40104T & FOS  & G190H & \hbox{}\hspace{2mm}900 &                    1573&2330 &                    1&41 & 93-11-18 \\
Y1C40103P & FOS  & G270H & \hbox{}\hspace{2mm}600 &                    2221&3300 &                    2&01 & 93-11-18 \\
Z1BN0106T & GHRS & G160M &                 3\,917 &                    1223&1259 &                    0&07 & 93-05-15 \\
Z1BN0107T & GHRS & G160M &                 3\,699 &                    1315&1351 &                    0&07 & 93-05-15 \\
\noalign{\smallskip}
\hline
\end{tabular}
\end{table*}

The high-resolution GHRS spectra allow to measure the radial velocity of \object{K\,648} precisely:
In Z1BN0106T and Z1BN0107T \sT{stecf} the photospheric lines
are blueshifted by 0.55 and 0.57\AA\, (we used the \Ion{C}{3}
\Ion{C}{4}, \Ion{N}{5}, and \Ion{O}{4} lines in the spectra to measure the shift) and hence
$v_\mathrm{rad} = -128$ and $-133\,\mathrm{km/sec}$, respectively. This is close to the values given by Joy
(1949, $-115$ and $-129\,\mathrm{km/sec}$) and Schneider \ea\, (1983, $-128\,\mathrm{km/sec}$). However, the velocity of
\object{M\,15} is somewhat smaller ($v_\mathrm{rad} = -107\,\mathrm{km/sec}$, Harris 1996).

The analysis of UV spectra requires a careful determination of the stellar continuum.
Thus we start already here with an investigation of the interstellar \Ion{H}{1} column density \sK{red} 
and the interstellar reddening \sK{red}. 
The NLTE spectral analysis is then described in \se{SpAn}.

\subsubsection{Interstellar sulfur lines}
\label{sis}

Two interstellar lines, \Ionww{S}{2}{1250.59, 1253.81},
are prominent in the GHRS spectra. This allows to measure the wavelength shift of the
ISM: It is blueshifted by about 0.145\AA\ (this is equivalent to $v_\mathrm{rel} = -35 \mathrm{km/sec}$). 
This values allows e.g.\, to identify the interstellar contribution to the photospheric
\Ion{N}{5} lines \sK{abund}.

\subsubsection{Interstellar neutral hydrogen and reddening}
\label{red}

A \Ion{H}{1} column density of $n_{\mathrm{H\,I}} =  5(\spm 1)\cdot 10^{20} \mathrm{cm^{-2}}$ 
is determined from the Ly\,$\alpha$ 
profile \sA{fnh}.
The interstellar reddening is then estimated by means of
the relation $n_{\mathrm{H\,I}} = (3.8\spm 0.9)\cdot 10^{21} E_\mathrm{B-V}$
(Groenewegen \& Lamers 1989) yielding a color excess of $E_\mathrm{B-V} = 0.13^{+0.08}_{-0.04}$.
We use this result to compare a synthetic spectrum (from a H+He model) with the FOS spectrum of 
\object{K\,648} \sA{ebv}. In agreement with the above values, we achieve a good fit at 
$E_\mathrm{B-V} = 0.10$ in the wavelength range 2100-3200\,\AA. This is also the amount of the
foreground reddening towards \object{M\,15} (Harris 1996).

\begin{figure}[ht]
  \centering
  \includegraphics[width=\hsize]{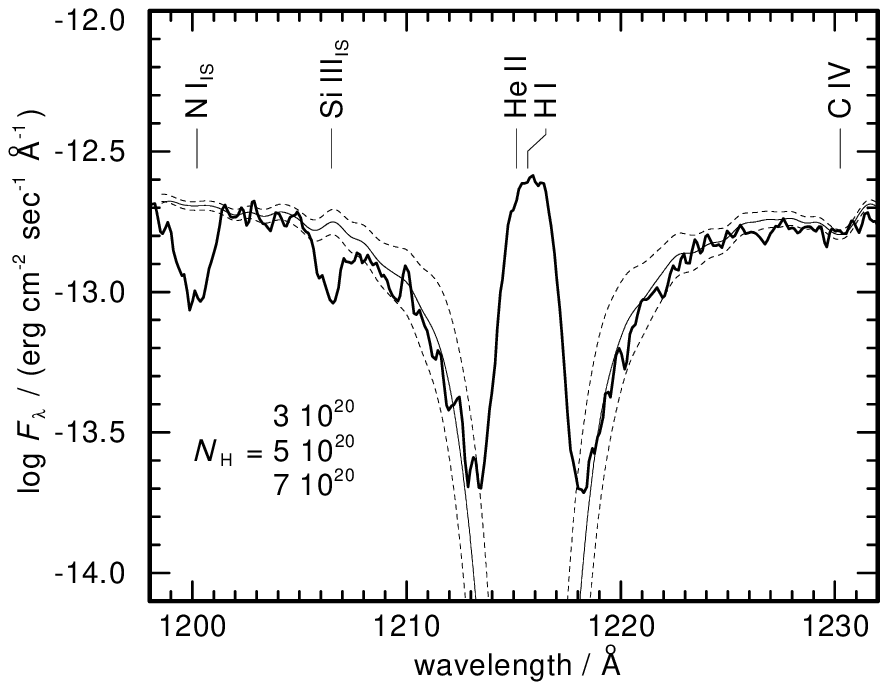}
  \caption[]{Comparison of the theoretical Ly\,$\alpha$ profile with the FOS spectrum. The best fit is
             achieved at $n_{\mathrm{H\,I}} =  5\cdot 10^{20}/\mathrm{cm^2}$. The theoretical flux is reddened according 
             to $n_{\mathrm{H\,I}}$ (the continuum is scaled to fit the line wings of Ly\,$\alpha$, \se{red}). 
             The parameters of the line-blanketed model atmosphere are summarized in \ta{php}
            }
  \label{fnh}
\end{figure}

At shorter wavelengths the model predicts a higher flux level than observed. This discrepancy is
not caused by metal-line blanketing. Even when we increase the abundances
to solar, no match can be achieved \sA{ebvi}. Instead there a significantly different reddening law 
might be valid.

Since IUE spectra (e.g.\,SWP\,17069)
show the same flux level like the FOS spectra in the wavelength range around 1400\AA\ (not shown), 
problems in the flux calibration appear to be unlikely.

\begin{figure}[ht]
  \centering
  \includegraphics[width=\hsize]{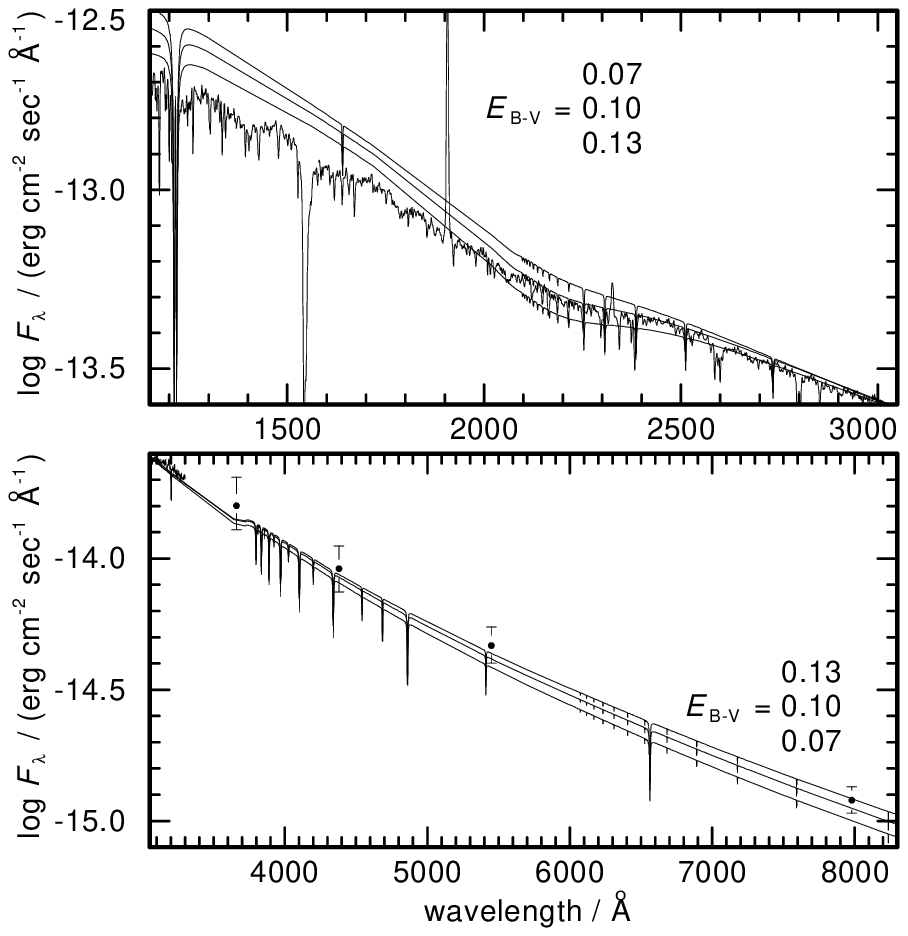}
  \caption[]{Determination of the interstellar reddening. A synthetic spectrum, calculated from a
             H+He model (\Teffw{39}, \loggw{3.9}, \ratiow{He}{H}{0.08}), 
             is shown with a reddening according to $E_\mathrm{B-V}$ = 0.07, 0.10, and 0.13.
             At longer wavelengths ($\lambda \sga 2000 \mathrm{\AA}$) the observation is well fitted
             with $E_\mathrm{B-V} = 0.10$ while at shorter wavelengths the reddening is underestimated.
             In the optical wavelength range (bottom), the measurements by Alves \ea (2000) are
             indicated with their error bars. The higher $E_\mathrm{B-V}$ seems to fit slightly better in
             the optical, however, then the fit to the measured UV flux is worse
            }
  \label{ebv}
\end{figure}

\begin{figure}[ht]
  \centering
  \includegraphics[width=\hsize]{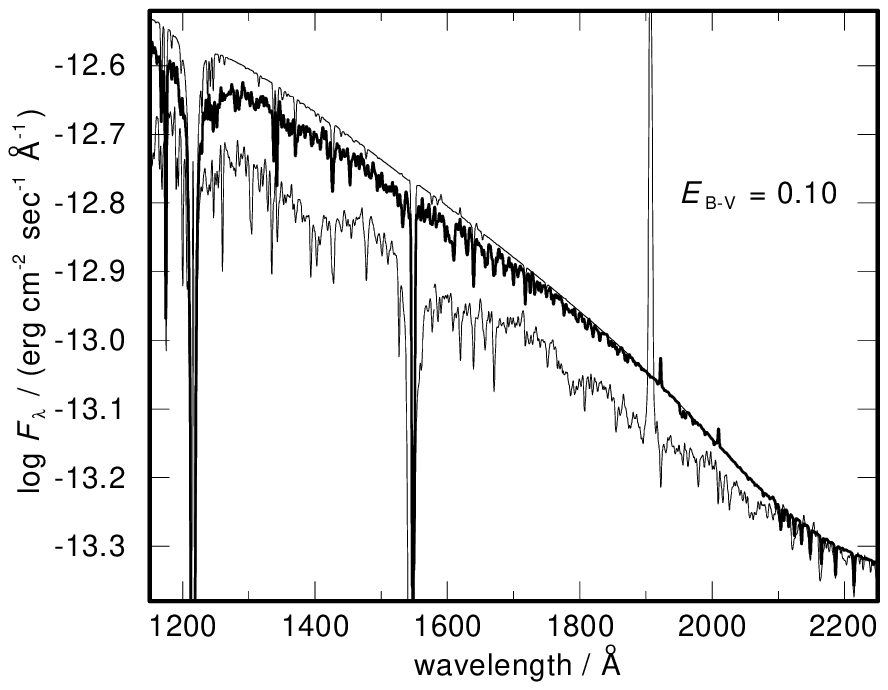}
  \caption[]{Impact of iron-group opacities on the UV flux (model parameters: \ta{php}). 
             Even at solar iron-group abundances (thick line),
             the model flux is much higher than observed. At the cluster metallicity [Fe/H] = -2.25, 
             the iron-group opacities can be almost neglected (fine line). The synthetic spectrum
             is scaled to fit the observed flux at 3\,000\,\AA}
  \label{ebvi}
\end{figure}

In the FOS spectra, we will use the \Ion{He}{2} Fowler series for our analysis \sA{fos} which is 
located in the part of the spectrum that is well matched by the model. 
Thus we adopt $E_\mathrm{B-V} = 0.10$ for our analysis.
In the case of detailed line profile fits to the FOS and GHRS spectra, 
the model continuum has to be individually scaled in order to fit the observed continuum around
the analyzed line.

\subsection{Optical spectra}
\label{opt}

The optical spectra (resolution = 1\,\AA) of \object{K\,648} were obtained at the 3.5m telescope
at the Calar Alto (DSAZ, Spain) on June 12, 1989 with an exposure time of 2100\,sec. 
Since these spectra are, in contrast to the UV spectra which are used here, not absolutely calibrated we use
the rectified spectrum (e.g.\,\ab{twin}) instead.

Since \object{K\,648} lies close to the center of
\object{M\,15}, the background determination was difficult (Heber \ea\,1993) due to
the low spatial resolution of the TWIN spectrograph.

Some of the ``strategic'' spectral lines are contaminated by nebular emission \sA{twin} and thus unreliable:
E.g. the \Ion{He}{1}\,/\,\Ion{He}{2} ionization equilibrium which is usually a very sensitive
indicator for the effective temperature \Teff\ cannot be evaluated because all \Ion{He}{1} lines
are filled in by nebular emission. However, \Ion{He}{2} lines as well as some carbon lines can be measured and
will be used in our spectral analysis.

\begin{figure*}[htb]
  \centering
  \includegraphics[width=\hsize]{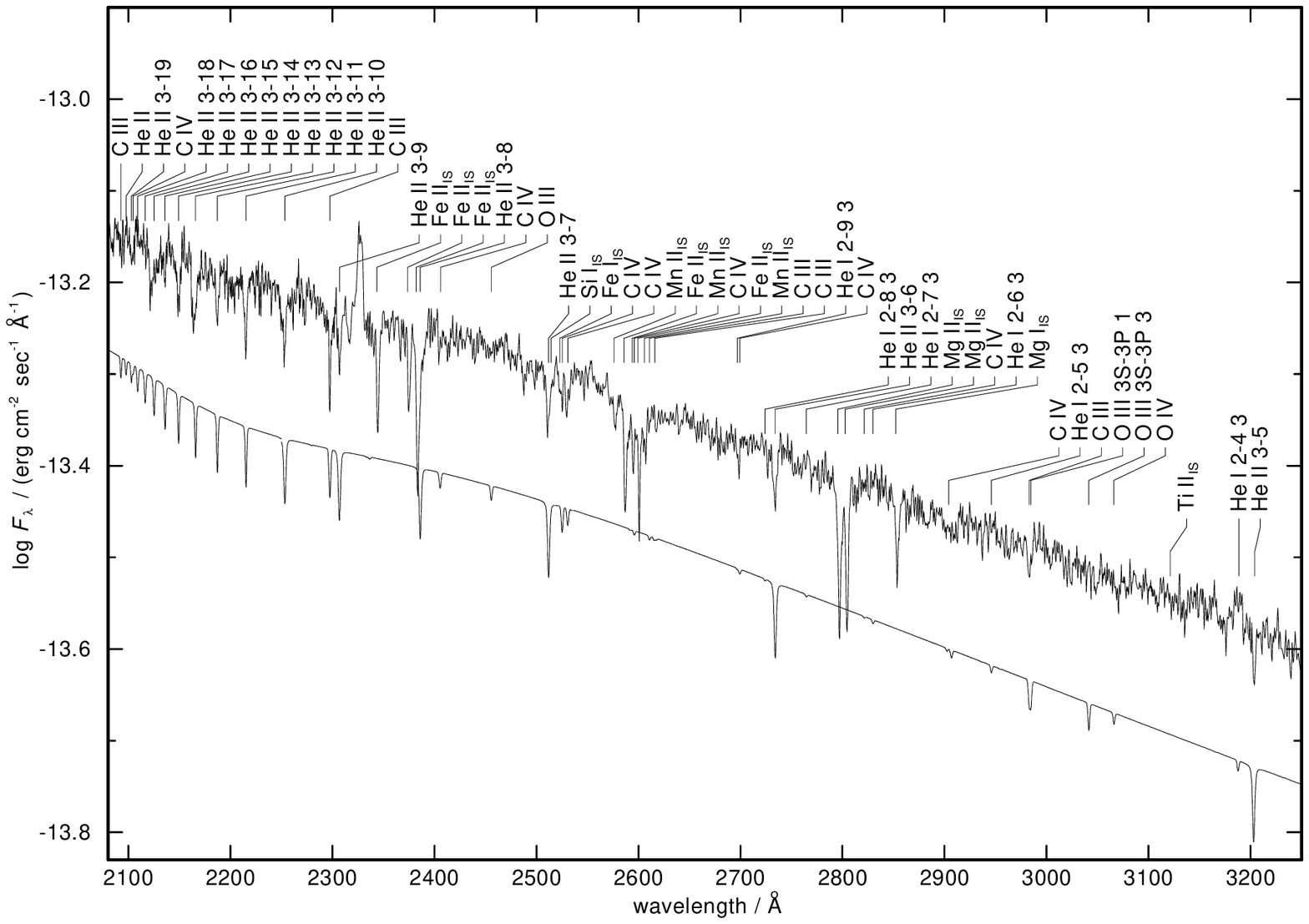}
  \caption[]{UV spectrum of K\,648 obtained with the FOS (top). Same like \ab{twin}. Note that the
             \Ion{He}{2} lines, calculated with \ratiow{He}{H}{0.08} are well matched
            }
  \label{fos}
\end{figure*}

\section{NLTE analysis}
\label{SpAn}

Bianchi \ea\,(1995) found evidence for a stellar wind of \object{K\,648} and derived
$v_\mathrm{\inf} = 1630 \mathrm{km/sec}$ and $\log \dot M / {\rm (M_\odot yr^{-1})} = -8.2$
from the strong \Ionww{C}{4}{1548, 1550} resonance doublet. Other spectral lines appear
unaffected by the wind. Hence, the plane-parallel, hydrostatic NLTE model atmospheres 
(Werner 1986, Rauch 1997, Rauch 2000, and references therein) are appropriate for this analysis.

In order to model the background opacity, metal-line blanketing of the iron-group elements 
(Dreizler \& Werner 1993, Haas \ea\,1996) is considered 
with a generic model atom. This is constructed from all elements Ca -- Ni (ionization stages {\sc iii -- vii} with
all lines given by Kurucz 1996). Reduced abundances ([Z/H] = -2.25, following Harris 1996) are assumed.
However, the impact of the Ca-Ni lines on the spectrum is small \sA{ebvi}. 

In order to check the validity of the preliminary parameters from Heber \ea\,{1993} 
(\Teffw{37}, \loggw{4.0}, \ratiow{He}{H}{0.5}) we have performed some test calculations. From the
comparison of synthetic spectra calculated from the new H+He models to the observation
we find that $g$ is in good agreement, while \Teff\ is higher and \ratio{He}{H} is
about solar. The differences can be explained by the fact that our NLTE model atmospheres have 
been significantly improved since 1993 (e.g.\, Rauch 1997, Rauch 2000): The metal-line blanketing of
all elements up to the iron group is considered in detail and thus the temperature structure is 
correctly modelled and subsequent line-formation calculations in order to determine metal
abundances are much more reliable. E.g.\, the high He abundance found by Heber \ea\,(1993)
and Bianchi \ea\,(2000) is due to artificial effects if model atoms are too small and/or
metal-line blanketing is not accounted for in the model atmosphere calculation.

Due to the impact of CNO on the atmospheric structure and background
opacity we decided to calculate a grid of H+He+C+N+O+(Ca-Ni) models in order to
use e.g.\, the \Ion{C}{3}\,/\,\Ion{C}{4} ionization equilibrium to determine \Teff\
reliably. The analysis is described in the following.

\subsection{Surface gravity and helium abundance}
\label{logg}

The first parameter to be fixed is $g$. For this purpose we use two strongest uncontaminated lines in the optical
spectrum, \Ionww{He}{2}{4200, 4541} \sA{twin}. We have calculated H+He models with \Teffw{37-43}, \loggw{3.5-4.5} at different
\ratio{He}{H} ratios. Although \Teff\ cannot reliably be judged from this approach, the \ratio{He}{H} is
about solar. \Ionw{He}{2}{4686} may be contaminated by nebular emission is not considered.
Models with \loggw{3.7-4.2} and \ratiow{He}{H}{0.06-0.10} (by number) fit the observation well \sA{ghe}.
These values are consistent with our more detailed models which account for metal-line blanketing (see below). We 
adopt \loggw{3.9} and \ratiow{He}{H}{0.08}.
An error of 0.3\,dex can be estimated by the variation of \Teff\ and \logg\ within their error range 
and the quality of the available spectrum.

\begin{figure}[ht]
  \centering
  \includegraphics[width=\hsize]{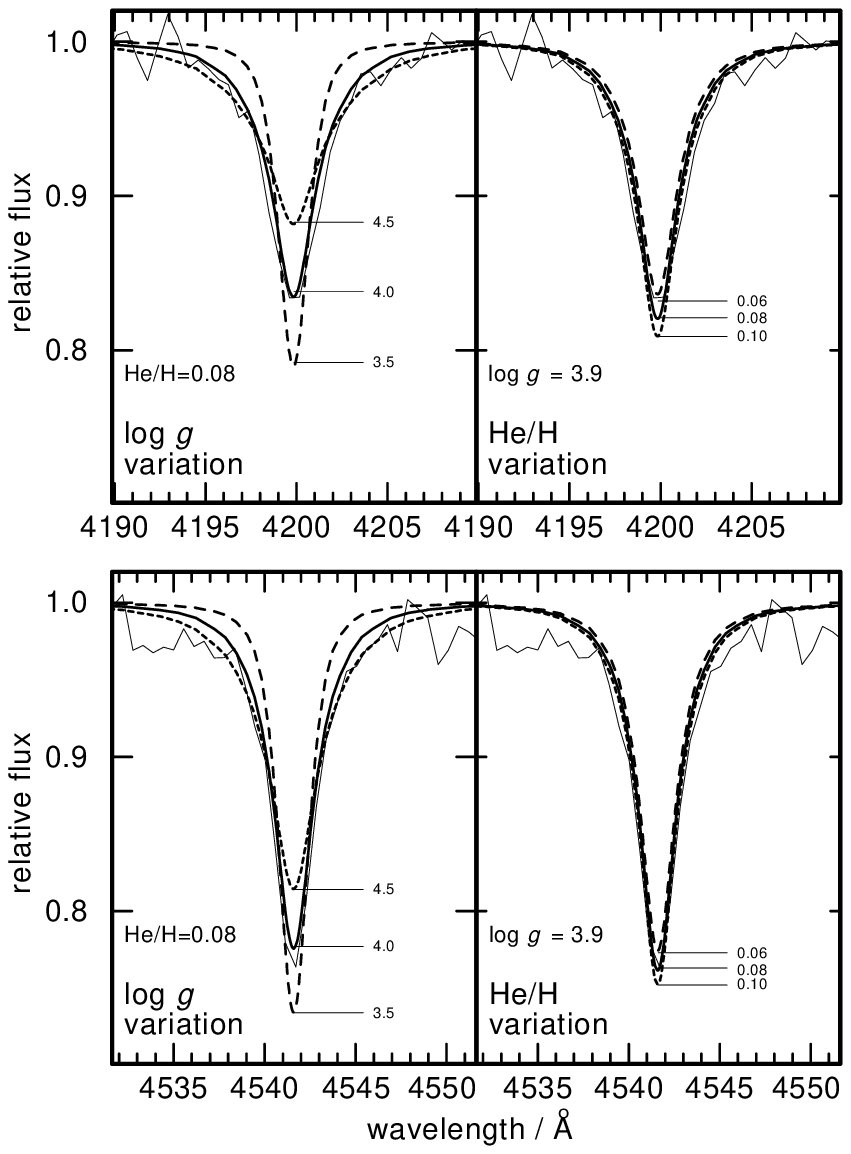}
  \caption[]{Theoretical line profiles (\Teffw{39}, convolved with a Gaussian of 1\AA\ FWMH in order to match the 
             instrumental resolution) of \Ionw{He}{2}{4200} (top) and \Ionw{He}{2}{4541} (bottom) 
             compared with the observation.
             At $3.7\sla$\logg\ or \logg$\sga 4.2$ no good fit can be achieved
             }
  \label{ghe}
\end{figure}

\subsection{Effective temperature and carbon abundance}
\label{teff}

In the optical spectrum of \object{K\,648}, the \Ion{He}{1} lines are
contaminated by nebular emission, and thus, a commonly used tool to determine
\Teff, the \Ion{He}{1}\,/\,\Ion{He}{2} ionization equilibrium, cannot be used.
The GHRS UV observations \sT{stecf} had been used with the aim to evaluate
the \Ion{C}{3}\,/\,\Ion{C}{4} ionization equilibrium (Heber \ea\,1993) which is also a very
sensitive indicator for \Teff (cf.\,Rauch 1993). 

Since there are a lot of carbon lines identified in the optical spectrum \sA{twin} we will use them to
determine the carbon abundance first and then model the \Ion{C}{3}\,/\,\Ion{C}{4} ionization equilibrium precisely.
New H+He+C+N+O+(Ca-Ni) models (\Teffw{39}, \loggw{3.9}, \ratiow{He}{H}{0.08}, [(Ca-Ni)/H] = -2.25 have been calculated and 
the CNO abundances were adjusted to fit the observation \sA{chop}. We adopt 
\ratiow{C}{H}{0.001} $\spm 0.3\,\mathrm{dex}$.

\begin{figure}[ht]
  \centering
  \includegraphics[width=\hsize]{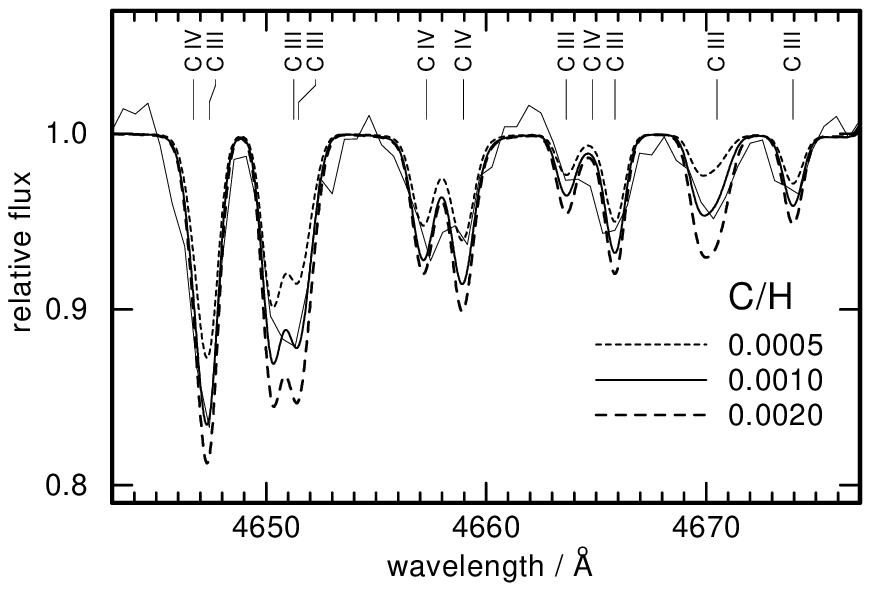}
  \caption[]{Synthetic spectra from models with \Teffw{39}, \loggw{3.9}, \ratiow{He}{H}{0.08} and
             different \ratio{C}{H} ratios compared with the observation.
             At \ratiow{C}{H}{0.0005} (about solar ratio) all C lines appear too shallow.
             At \ratiow{C}{H}{0.001} the \Ion{C}{3} lines are still too weak while the 
             \Ion{C}{4} are already too strong. This indicates a slightly lower \Teff. At higher
             \ratio{C}{H} ratios, no fit of the C lines is possible 
             }
  \label{chop}
\end{figure}

We calculated synthetic spectra from a small grid of models with \Teffw{37-43} and \ratiow{C}{H}{0.001}.
It turned out that the \Ion{C}{3} lines are strongly dependent on variation of \Teff\ while the 
\Ion{C}{4} lines appear almost unchanged \sA{fos1}. From the fit of the
strongest \Ion{C}{3} line in the FOS spectrum of \object{K\,648}, the \Ionw{C}{3}{1175} triplet, we determine
\Teffw{39\spm 2} \sA{fos1}. 

\begin{figure}[ht]
  \centering
  \includegraphics[width=\hsize]{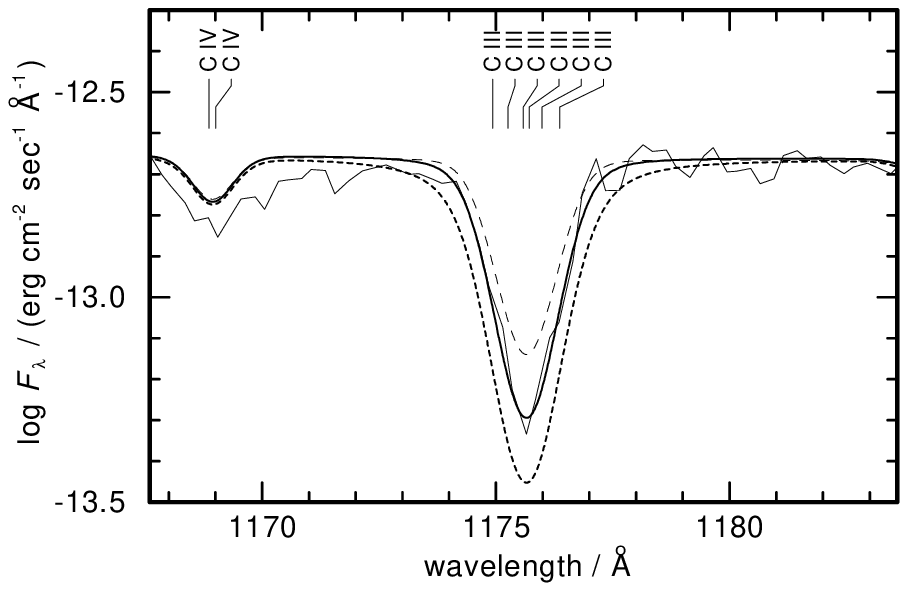}
  \caption[]{Synthetic spectra with \Teff\,= 37 (short dashes), 39 (fully drawn), and 41 (long dashes) kK
             compared with the FOS spectrum \sT{stecf}.
             At \Teffw{39} \Ionw{C}{3}{1175} is well matched
             }
  \label{fos1}
\end{figure}

This result is verified by the \Ion{C}{3} and \Ion{C}{4} lines in the high-resolution GHRS spectra \sA{hrs1}.

\begin{figure}[ht]
  \centering
  \includegraphics[width=\hsize]{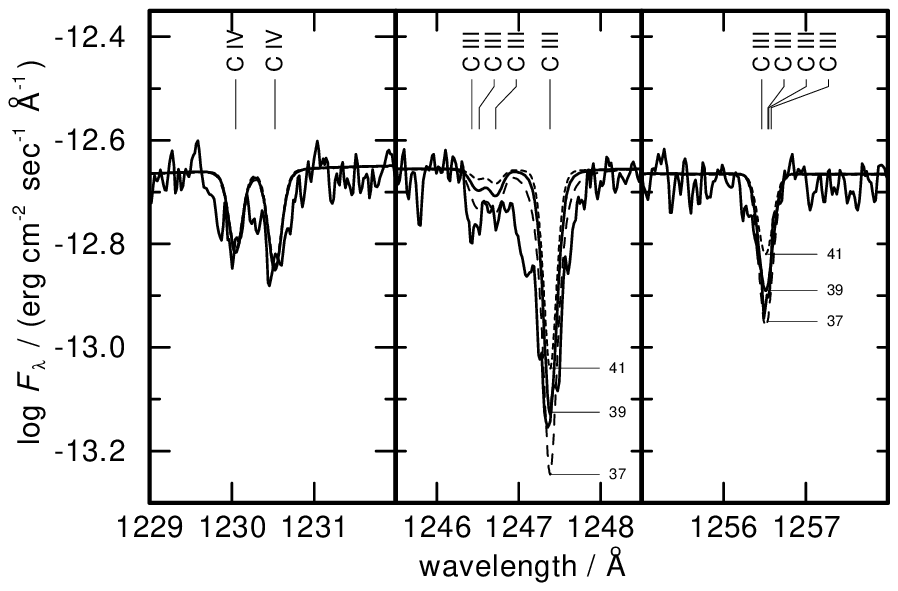}
  \caption[]{Synthetic spectra with \Teff\,= 37 (long dashes), 39 (fully drawn), and 41 (short dashes) kK
             compared with the GHRS spectrum \sT{stecf}.
             \Ionw{C}{3}{1247} is reproduced at \Teffw{39}, \Ionw{C}{3}{1256} at \Teffw{38}.
             \Ionw{C}{4}{1230} is almost unchanged in this \Teff\ range
             }
  \label{hrs1}
\end{figure}

\subsection{Photospheric nitrogen and oxygen abundances}
\label{abund}

In the last step, we determine the N and O abundances. New H+He+C+N+O + iron-group \sK{SpAn} models 
are calculated. In the GHRS spectra, the most
prominent N and O lines are the \Ionw{N}{5}{1238, 1242} resonance doublet (apparently unaffected by the
stellar wind \sK{SpAn} and the \Ionww{O}{4}{1338, 1343}
lines. We use these in order to determine the abundances. We achieve \ratiow{N}{H}{1\cdot 10^{-6}}
\sA{n2h} and \ratiow{O}{H}{1\cdot 10^{-3}} \sA{o2h}. 
Due to the S/N ratio of the spectra, an error of $\spm 0.5\,\mathrm{dex}$ has to be assumed.
Note that the line cores are much too strong
at higher abundances.

\begin{figure}[ht]
  \centering
  \includegraphics[width=\hsize]{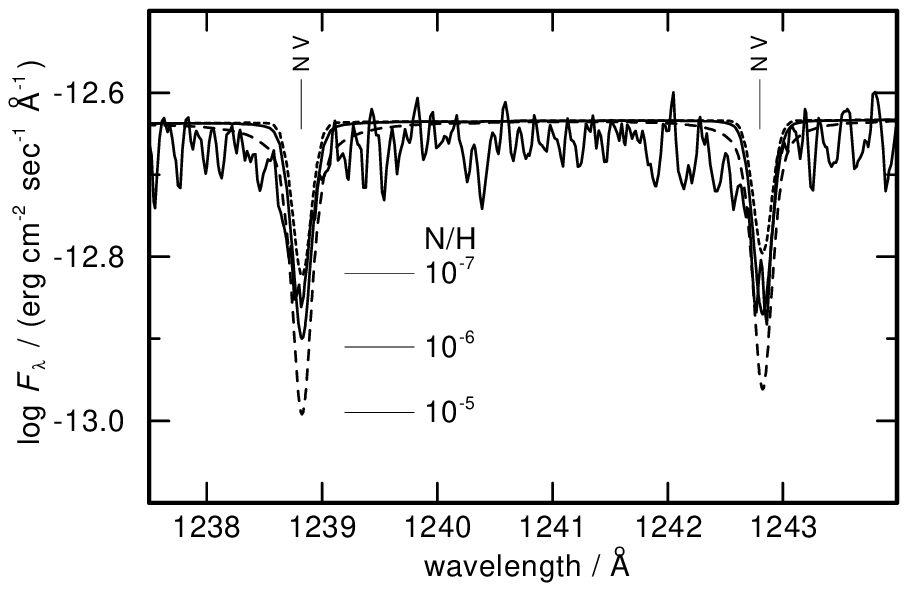}
  \caption[]{Synthetic spectra with \Teffw{39}, \loggw{3.9}, \ratiow{He}{H}{0.08},
             \ratiow{C}{H}{1\cdot 10^{-3}}, \ratiow{O}{H}{1\cdot 10^{-3}}, and different N
             compared with the GHRS spectrum \sT{stecf}.
             \Ionw{N}{5}{1238, 1242} is reproduced at \ratiow{N}{H}{1\cdot 10^{-6}}
             }
  \label{n2h}
\end{figure}

\begin{figure}[ht]
  \centering
  \includegraphics[width=\hsize]{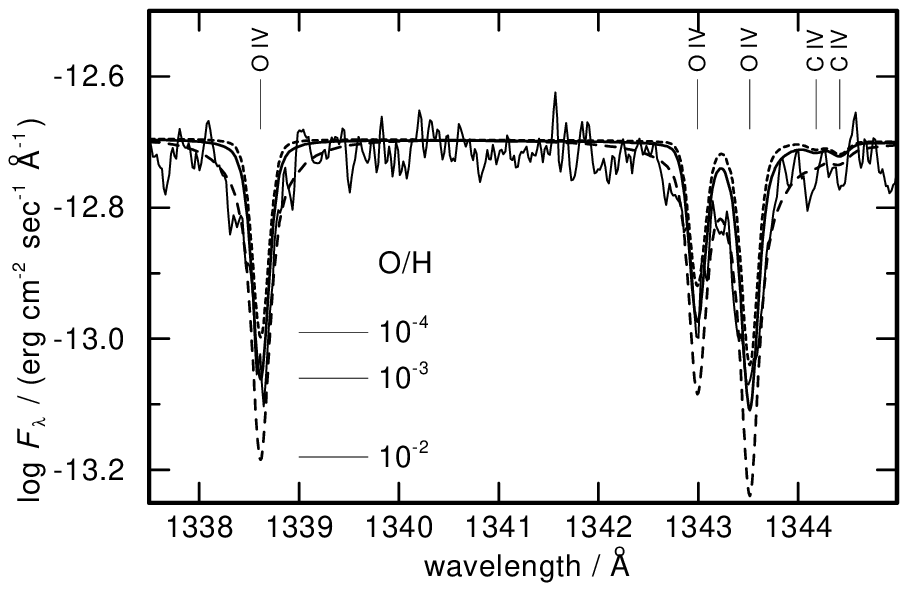}
  \caption[]{Synthetic spectra with \Teffw{39}, \loggw{3.9}, \ratiow{He}{H}{0.08},
             \ratiow{C}{H}{1\cdot 10^{-3}}, \ratiow{N}{H}{1\cdot 10^{-5}}, and different O abundances
             compared with the GHRS spectrum \sT{stecf}.
             \Ionww{O}{4}{1338, 1343} is reproduced at \ratiow{O}{H}{1\cdot 10^{-3}}
             }
  \label{o2h}
\end{figure}

\section{Mass, luminosity, post-AGB age, and spectroscopic distance}
\label{MLAD}

In \ab{evo} we compare the position of \object{K\,648} with standard evolutionary
tracks of hydrogen-burning post-AGB stars. From the evolutionary calculations
of Sch\"onberner (1983) we interpolate a stellar mass of $0.57^{+0.02}_{-0.01}\,\mathrm{M}_\odot$,
a luminosity of $3\,810\spm 1200\,\mathrm{L}_\odot$, and a post-AGB age of $6\,800^{+3\,500}_{-2\,100}$\,a.

\begin{figure}[ht]
  \centering
  \includegraphics[width=\hsize]{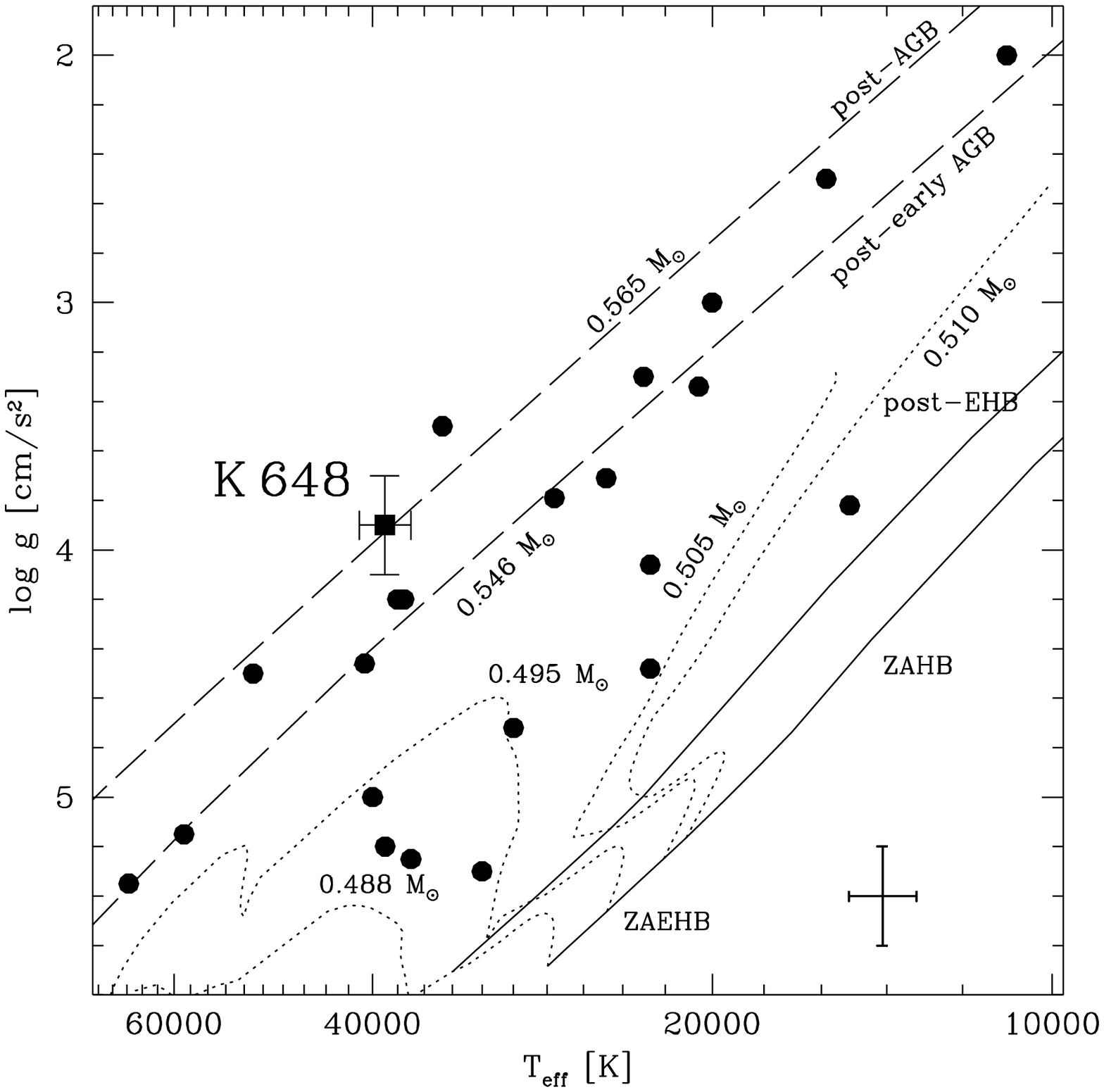}
  \caption[]{Position of \object{K\,648} in the $\log \Teff$--$\log g$ plane
             compared to theoretical evolutionary tracks of post-AGB
             hydrogen-burning stars (dashed, masses given in
             $\mathrm{M_\odot}$, Sch\"onberner 1983) and post-EHB evolutionary tracks for [Fe/H] = -1.48
             (Dorman \ea\,1993).
             The positions of other UV-bright stars in globular clusters (cf.\,Moehler \ea\,1998, Fig.\,3)
             are marked. The cross in the lower right part indicates the typical errors 
             }
  \label{evo}
\end{figure}

From our final model \sT{php}, we can determine the spectroscopic distance of \object{K\,648}
using the flux calibration of Heber \ea\,(1984):

\begin{equation}
f_{\rm V} = 3.58\cdot 10^{-9}\cdot 10^{{\rm -0.4}m_\mathrm{V_0}} \mathrm{erg\,cm^{-2}\,sec^{-1}\,\AA\hbox{}^{-1}}
\end{equation}

\noindent
with $m_\mathrm{V_0} = m_\mathrm{V} - 2.175 c$, $c = 1.47 E_\mathrm{B-V} = 0.147$ \sA{ebv},
$m_\mathrm{V} = 14.73$, and $M = 0.57 {\rm M_\odot}$, the distance is derived from

\begin{equation}
d = 7.11\cdot 10^{4} \sqrt{H_\nu\cdot M\cdot 10^{0.4m_\mathrm{V_0}-\log g}}~{\rm pc}\,.
\end{equation}

\noindent
With the Eddington flux at $\lambda_{\rm eff} = 5454\ang$ our final model atmosphere
$H_\nu = 5.31\cdot 10^{-4} {\rm erg\,cm^{-2}\,sec^{-1}\,Hz^{-1}}$
we derive a distance of $d= 11.1^{+2.4}_{-2.9}\,\mathrm{kpc}$.
This distance is in agreement with the distance of the globular cluster \object{M\,15}
($d = 12.3\spm 0.6\,\mathrm{kpc}$, Alves \ea\, 2000).

\section{The surrounding nebula \object{Ps\,1}}
\label{ps1}

For the ambient nebula, we can calculate the following: 
The linear dimension of the inner shell
($0\farcs 8\times 0\farcs 6$, Alves \ea\,2000) is    $0.043^{+0.009}_{-0.011} \times 0.032^{+0.007}_{-0.008} \,\mathrm{pc}$,
and of the outer shell
($3\farcs 1\times 2\farcs 7$, Alves \ea\,2000) it is $0.166^{+0.035}_{-0.044} \times 0.145^{+0.031}_{-0.038} \,\mathrm{pc}$.
If we assume an average expansion velocity of $v_\mathrm{exp} = 20\,\mathrm{km/sec}$, then the
dynamical age is $t_\mathrm{dyn} = 3950^{+\,\,850}_{-1050}\,\mathrm{a}$. This is within the error range in
agreement with its post-AGB age ($6\,800^{+3\,500}_{-2\,100}$\,a, \se{MLAD}).

\section{Results and discussion}
\label{dis}

In our analysis it became clear that the determination of \Teff\ from the two available GHRS spectra alone 
(\Ion{C}{3}\,/\,\Ion{C}{4} ionization equilibrium) is not precise enough (too few lines to evaluate) -- together
with the available IUE, FOS, and TWIN spectra, however, we determined \Teffw{39\spm 2} and \loggw{3.9\spm 0.2} which is 
within the error ranges in good agreement with the preliminary values given by Heber \ea\,(1993, $37\,\mathrm{kK}\,/\,4.0$). 
In contrast to Heber \ea, however, we arrive at an almost solar (\ratiow{He}{H}{0.08}) He abundance. The high
carbon abundance (2.5 times solar) is verified. The photospheric parameters of \object{K\,648} are summarized in \ta{php}.

\begin{table}[ht]
\caption[]{Parameters of \object{K\,648}. The errors are estimated by the variation of
           \Teff, \logg\ and the abundance ratios within their error limits in order to achieve a fit of the synthetic
           spectrum to the observation. 
           The solar abundance ratios are given by Holweger (1979) and St\"urenburg \& Holweger (1990).
           The cluster metallicity is $\frac{1}{178}$ times solar (Harris 1996)} 
\label{php}
\begin{tabular}{r@{\,/\,}lr@{.}llr}
\Teff           & \kK                         &  39&0                         & \spm 2.0     \\
$\log (g$& $\mathrm{\frac{cm}{s^2}})$         &   3&9                         & \spm 0.2     & \mc{1}{r}{solar value}\\
\cline{6-6}
\noalign{\smallskip}
$n_\mathrm{He}$       & $n_\mathrm{H}$              & \mc{2}{r}{$8\cdot 10^{-2}$  } & \spm 0.3 dex & $1\cdot 10^{-1}$ \\
$n_\mathrm{C}$        & $n_\mathrm{H}$              & \mc{2}{r}{$1\cdot 10^{-3}$  } & \spm 0.3 dex & $4\cdot 10^{-4}$ \\
$n_\mathrm{N}$        & $n_\mathrm{H}$              & \mc{2}{r}{$1\cdot 10^{-6}$  } & \spm 0.5 dex & $1\cdot 10^{-4}$ \\
$n_\mathrm{O}$        & $n_\mathrm{H}$              & \mc{2}{r}{$1\cdot 10^{-3}$  } & \spm 0.5 dex & $7\cdot 10^{-4}$ \\
\hline
\noalign{\smallskip}
$M$                   & $\mathrm{M_\odot}$          &   0&57                        & $^{+0.02}_{-0.01}$ \\
$L$                   & $\mathrm{L_\odot}$          & \mc{2}{c}{3\,810}             & \spm 1200 \\
$d$                   & $\mathrm{kpc}$              & \mc{2}{c}{11.1}               & $^{+2.4}_{-2.9}$ \\
$t_\mathrm{post-AGB}$ & $\mathrm{a}$                & \mc{2}{c}{6\,800}             & $^{+3\,500}_{-2\,100}$ \\
\noalign{\smallskip}
\hline
\noalign{\smallskip}
$t_\mathrm{dyn(PN)}$  & $\mathrm{a}$                & \mc{2}{c}{3\,925}             & $^{+830}_{-1036}$ \\
\noalign{\smallskip}
\hline
\end{tabular}
\end{table}

The solar \ratio{He}{H} surface abundance indicates that \object{K\,648} might be a H-burning
post-AGB star. Compared to the metallicity of \object{M\,15}, the C and O abundances appear
enriched \sA{x2h}, likely due to a He-shell flash. 

\begin{figure}[ht]
  \centering
  \includegraphics[width=\hsize]{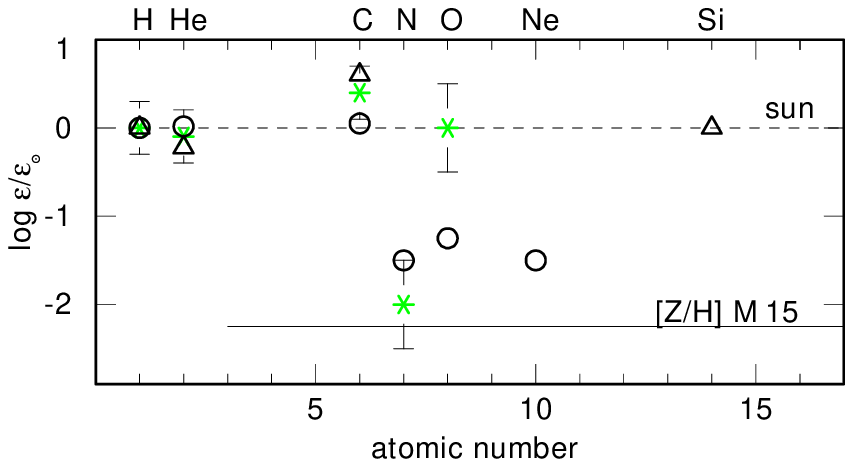}
  \caption[]{Photospheric abundances of \object{K\,648} 
            ($\ast$: this work, $\triangle$: Bianchi \ea\,2001) and its PN ($\circ$: Adams \ea\,1984) 
             relative to the sun. The metalicities of the sun and of M\,15 are indicated
            }
  \label{x2h}
\end{figure}

In order to further improve the analysis and the abundance determinations, 
high-resolution and high-S/N spectra with high spatial resolution
(minimize the contamination by the nebula) are highly desirable:   
Such spectra which cover the complete UV range 
(many lines of different elements and different ionization stages $\rightarrow$
improvement of \Teff, cf.\,Rauch 1993)
will separate the possible interstellar and photospheric lines, 
e.g. in \Ionww{N}{5}{1238, 1242}. Moreover, we can identify \Ionww{Si}{4}{1393, 1402} and
\Ionw{S}{5}{1502} in the available FOS spectra at a low S/N ratio which hampers a reliable
analysis. Better spectra allow a precise abundance determination of these elements. 
The optical wavelength range (down to the Balmer edge $\rightarrow$ improvement of $g$)
gives additional constraints.

\begin{acknowledgements}
This research was supported by the DLR under grant 50\,OR\,9705\,5 (T\"ubingen)
and 50\,OR\,9602\,9-ZA (Bamberg).
Computations were carried out on CRAY computers of the Rechenzentrum der Universit\"at Kiel, Germany.
This research has made use of the SIMBAD Astronomical Database, operated at CDS, Strasbourg, France,
of the ST-ECF spectra database, and of the NIST Atomic Spectra Database.
\end{acknowledgements}

\end{document}